\documentclass{iau}
\usepackage{graphicx}

\title[Data-mining Based Expert Platform for the Spectral Inspection] 
{Data-mining Based Expert Platform for the Spectral Inspection}

\author[Hai-Jun Tian]   
{Hai-Jun Tian$^{1,2}$, Yang Tu$^{1}$,
Yan-Xia Zhang$^2$,
Yong-Heng Zhao$^2$,
Guo-Hong Lei$^1$,
Bo-Liang He$^2$,
Chen-Zhou Cui$^2$,
Xue-Lei Chen$^2$
}

\affiliation{$^1 $China Three Gorges University, Yichang, 443002. Email: {\tt hjtian@lamost.org}\\[\affilskip]
$^2$National Astronomical Observatories, Chinese Academy of Sciences, Beijing 100012}

\pubyear{2014}
\volume{306}  
\pagerange{119--126}
 \jname{Statistical Challenges in 21st Century
Cosmology} \editors{Alan Heavens, Jean-Luc Starck \& Alberto
Krone-Martins, eds.}
\begin{document}

\maketitle

\begin{abstract}
We propose and preliminarily implement a data-mining based platform
to assist experts to inspect the increasing amount of spectra with
low signal to noise ratio (SNR) generated by large sky surveys. The
platform includes three layers: data-mining layer, data-node layer
and expert layer. It is similar to the GalaxyZoo project and
VO-compatible. The preliminary experiment suggests that this
platform can play an effective role in managing the spectra and
assisting the experts to inspect a large number of spectra with low
SNR.


\end{abstract}

\section{Introduction}

With the telescopes established and the surveys ongoing, such as the
Guoshoujing telescope (LAMOST, \cite{Zhao99}), more and more
spectra are collected and released. Unfortunately, except the
qualified data, there still exist many spectra unclassified by the
automated pipeline. Usually most of these spectra are too low SNR to
be classified because of the limited magnitude of the instruments or
other reasons. Some important new discoveries are probably hidden in
these unknown spectra. Therefore, we should not give up these
seemingly useless data, even though they are quite defective. How to
handle these unknown spectra is one of the biggest challenges to the
modern statistics and data mining techniques as well as eyeball
check. In order to ensure the accuracy of the results, we have to
motivate experts to check these spectra by visual inspection. Owing
to huge amount of such spectra generated continuously by the large
sky surveys, it will spend much time and efforts to check spectra
one by one. Consequently, a platform to efficiently manage and
coarsely classify these unqualified data is in great requirement for
the large surveys.

\section{Architechure}\label{arch}
We propose such a platform, which aims to effectively process and
manage a large volume of spectra with low SNR, and mine as much as
possible knowledge from the spectra for the scientific research
through the integration of machine learning and human inspection.
The platform is a three-tier structure including data-node layer,
data mining layer and expert layer, as shown in Fig. 1. The core
part is data mining. Before the visual inspection, the dataset will
be preprocessed with various up-to-date statistical and data-mining
techniques. For instance, the dataset can be classified roughly as
stars, galaxies or quasars according to the limited information
contained in the low SNR spectra. All results from this layer could
efficiently help the expert to inspect the spectra. The expert layer
provides an interface by which anyone may query and inspect the
spectra and save the results in the data node layer. Finally the
system can give a more reliable result by efficiently utilizing the
experts' input (one spectrum may be inspected by more than one
expert). The platform will be VO-compatible and integrated
seamlessly with other surveys. Its client will be of diverse forms,
such as Matlab-GUI, Java-application, Java-applet, etc.

Several kinds of data-mining algorithms to pre-process the LAMOST
spectra have been investigated. For example, Support Vector Machines
(SVMs) is firstly applied to classify the LAMOST spectra, and the
classified spectra are saved into the data-node layer as a public
dataset, then the spectra may be queried with a web client and
inspected by a Java applet tool, finally the analysis results are
returned back into MyDB in the data-node layer.

\begin{figure}
\centering
\includegraphics[width=\linewidth, clip]{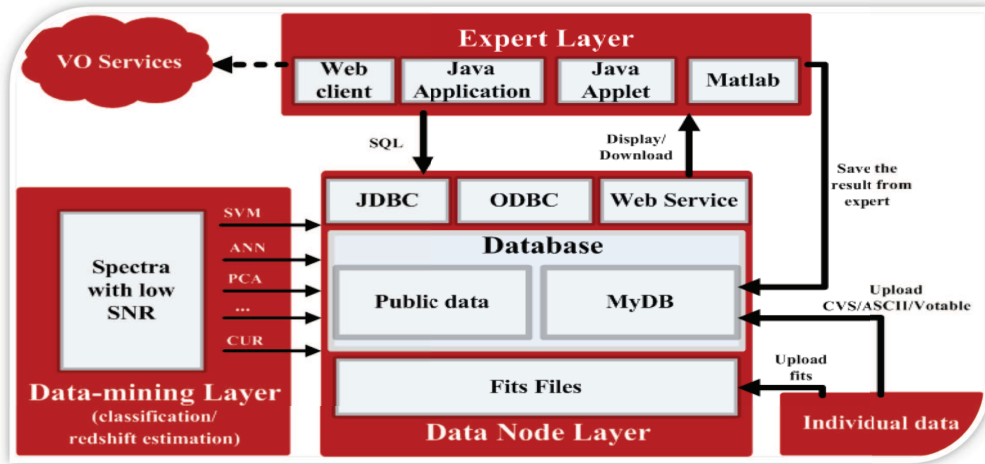}
\caption{Architecture of expert platform for the spectral inspection}
\label{architecture}
\end{figure}

{\underline{\it Data-mining layer}}. The data-mining layer has two
main tasks: classification and redshift or radial velocity
estimation for low SNR spectra using various advanced data-mining
techniques, such as Support Vector Machines (SVMs, \cite{PN_etal12}), Artificial
Neural Networks (ANN, \cite{zhang09}), K-Nearest Neighbors(KNN, \cite{li08}), Principal Component
Analysis(PCA, \cite{zhang03}; \cite{Yip04}), Wavelet(\cite{Machado_etal13}), CUR Matrix Decomposition(\cite{Yip14}).

{\underline{\it DataNode layer}}. The data-node layer includes two
parts: one is based on the database, which manages the catalog
generated by the data-mining layer (public dataset available for
each user) and the private dataset input by the users (MyDB, visible
only for the data owner); the other needs enough storage space to
save the spectral fits files, which are also divided into the public
and private. The results from the expert layer can also be saved in
MyDB as the private file. The data-node layer provides two
interfaces, e.g. the database interface (JDBC or ODBC) and the Web
Service interface, by which expert layer can query catalogs and
retrieve spectra.

{\underline{\it Expert layer}}. The clients with diverse forms, such
as Matlab-GUI, Java-application, Java-applet, etc., can be used to
display the spectra and assist the experts to inspect the spectra in
this layer. These tools provide all kinds of spectral templates and
reference lines, which can be easily shaped at expert's will. The
redshift can be calculated in real time as the template is moved.
The templates will automatically be best-matched based on the least
square method, and the optimal redshift can be figured out when the
spectrum is best fitted with the template.

The clients will support two kinds of queries: one is a simple way,
which creates SQLs through choosing the fields and filling the query
criteria; the other one provides a input form, which allows users to
compose much more complex SQL statements, and also provides the
functions of checking query syntax and database privilege. Through
the JDBC, web service or VO interfaces, the clients can access the
local or remote databases. Experts with different levels have
different weights to contribute the final results.

\begin{figure}
\centering
\includegraphics[width=0.49\linewidth, height=4cm, clip]{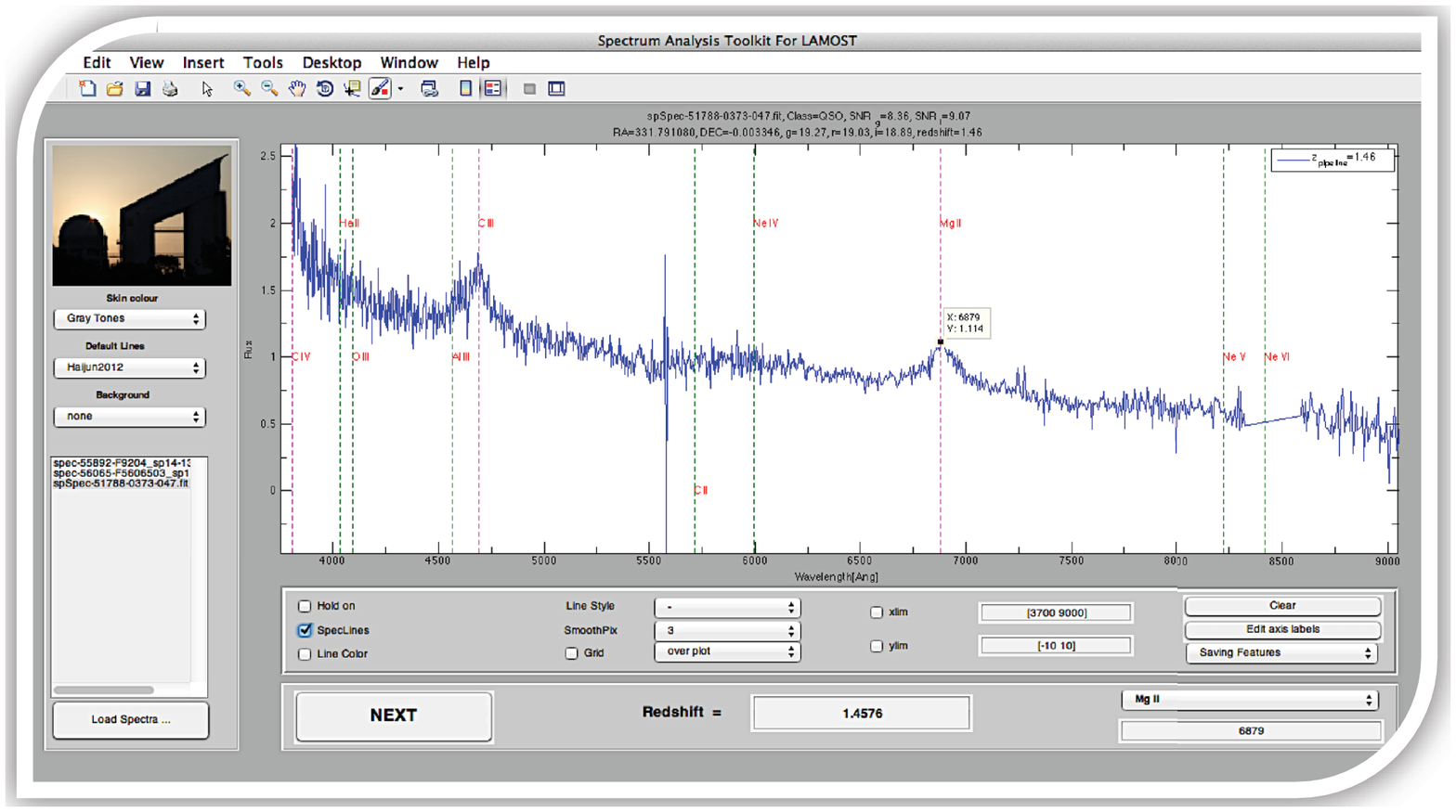}
\includegraphics[width=0.49\linewidth, height=4cm, clip]{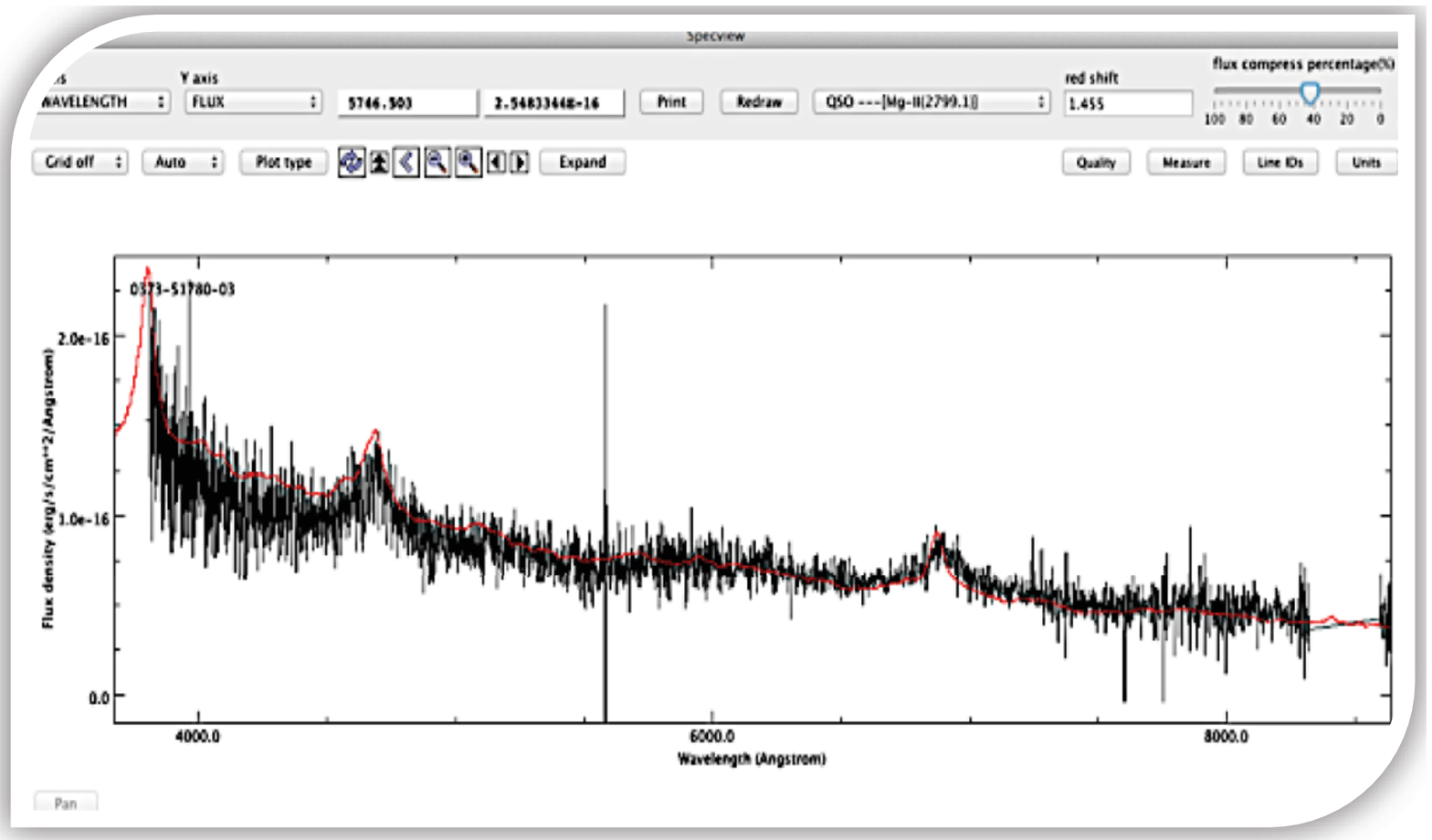}
\caption{Matlab (left) and Java (right) clients for the spectral inspection.}
\label{svm}
\end{figure}

For the data-node layer, we have finished the database and file
systems with public data and MyDB, and released the JDBC and Web
Service interfaces for the query functions.

For the expert layer, a Matlab client and a Jave client have been
preliminary developed, as displayed in the figure \ref{svm}. Actually, the Java client is modified SpecView, an interactive
java tool for visualization and analysis of spectral data
(\cite{Busko00}), in which we extended some functions including the
template fitting, automatic redshift estimation, the interaction
with the data-node layer, etc.

\section{Conclusions}
In order to overcome the difficulty in processing the poor spectra,
we propose and develop a data-mining based platform to encourage and assist
experts to inspect the low SNR spectra. With the help of the
platform, we tentatively process LAMOST spectra. The
preliminary experiment indicates that this platform indeed can play
an effective role in assisting experts to recognize a large number
of spectra with low SNR.

\section{Acknowledgments}
The authors thank the grants (No. U1231123, U1331202, U1331113, U1231108, 11073024, 11103027, 11303020) from NSFC. THJ thanks the support from LAMOST Fellowship.

\end{document}